\documentclass{camera}
\usepackage{epsfig} 
\def\bentarrow{\:\raisebox{1.3ex}{\rlap{$\vert$}}\!\longrightarrow}

\begin{document}
\renewcommand{\thefootnote}{\fnsymbol{footnote}}

\begin{flushright}
UPRF-2001-17\\
hep-ph/0107021\\
\end{flushright}

%
\title{THE LEP TRAIL TO NON-PERTURBATIVE~QCD$^*$}

%
\author{Matteo Cacciari}

%
\organization{
Dipartimento di Fisica, Universit\`a di Parma, and\\
INFN, Sezione di Milano, Gruppo Collegato di Parma}

\maketitle

\footnotetext[1]{Summary talk of
the Hadronic Physics session at ``LEPTRE'', XIII meeting on physics at LEP,
Rome, April 2001. To be published in the Proceedings.}
\renewcommand{\thefootnote}{\arabic{footnote}}

%
When giving a summary talk, one always faces a difficult choice.
Either she
makes a very personal selection of what to present, and integrates it
with more material, so as to produce a  coherent presentation (and
NOT give a summary at all!), or she sticks to some sort of accountant-like
report of the various parallel-session presentations (and she ends up with a
possibly dull talk).

I find the first alternative more attractive. However, on one hand it
doesn't do any justice to at least part of the speakers and, on the
other, it is difficult to prepare a proper talk in the few hours
available between the last parallel talk and the summary one
(not to mention the long-lasting social dinner and its excellent wine).

All in all, I have opted for what I hope to be a reasonable compromise:
reporting from all the speakers, but trying to frame the various
presentations into some kind of common path.

``{Path}'' being indeed the proper word, because the theme I have
chosen is the California gold rush in 1849.

The gold mine of present-day QCD is, of course, a full theoretical
calculation also including  what is today referred to as
``non-perturbative physics'' and usually parametrized and/or
fitted to data rather than derived from first principles.

After twelve years of running of LEP, and endless tests of QCD (or,
rather, of its perturbative regime) we are in a situation akin to the
one of the United States around mid-nineteenth century. The eastern
part, up to the Missouri-Mississippi river, has been colonized. Our
understanding of (and confidence in) perturbative QCD is on a very firm
ground. A few wild spots may of course still remain, but by far and
large this is now friendly territory.

However, far away, in California, we know the ultimate prize to lie in
wait: to be able to make theoretically sound predictions for what we now
generically refer to as non-perturbative physics.

\begin{figure}[t]
\begin{center}
\epsfig{file=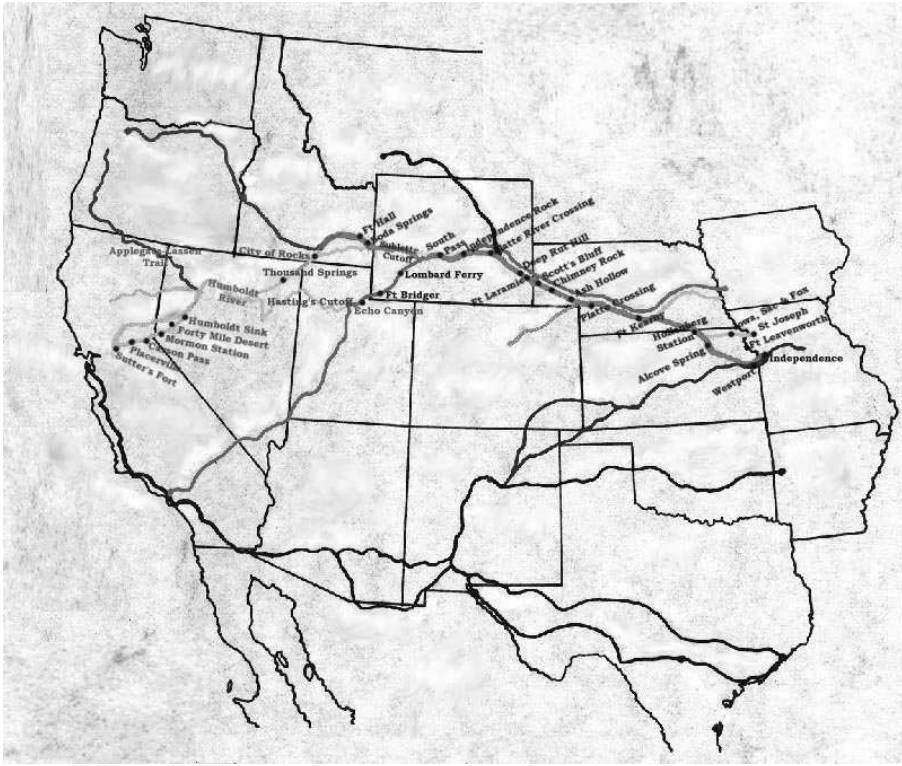,width=12cm}\hfill
\caption{\label{fig:trail}\small The California-Oregon trail, and its
alternatives, from the Missouri river to the West Coast.}
\end{center}
\end{figure}

``Young men attracted to California's golden opportunities in 1849 faced
two major problems. Finding gold was the second. The first was getting
there.''~\cite{web}
These young men endured a dangerous four to six month long
trip along the so-called California Trail (see Fig.~\ref{fig:trail}). The
ultimate goal was clear in their mind. How to get there much less.

It seems to me the path to understanding non-perturbative QCD might
live a similar situation. For the trip to California, more than a
single trail was traveled, as people tried different options, shortcuts
and cutoffs. In the same way, at least two regions of QCD are at
present being intensely studied in the intent of better understanding how to
reach the final goal. They are the power corrections and the
beyond-next-to-leading calculations, and they can be thought of as the 
North-West pass and the South-West pass route of the path to California.

Before describing in some detail the recent results obtained in these
two fields, I should however dedicate some time to more limited, but
nonetheless important, aspects which are being studied, and to some
experimental results obtained so far.

Let us start at the lower end of the energy spectrum. Graziano
Venanzoni~\cite{venanzoni}
reported on experimental results on $\sigma(e^+e^- \to {hadrons})$
at centre-of-mass energies below 5 GeV. The precise determination of
such a cross section has considerable importance for the physics at
much larger energies. In fact, it enters, via the so-called hadronic
contribution to the vacuum polarization, the evolution of the
electromagnetic coupling $\alpha_{em}(Q^2)$, and consequently the
precision physics at LEP1 and LEP2 energies.

Venanzoni also put forward the suggestion of using radiative events at
DA$\Phi$NE to measure the $e^+e^- \to {hadrons}$ cross section at
different centre-of-mass energies, rather than performing a machine-driven
energy scan as customary.

Remaining in the few-GeV energy region, Saverio
Braccini~\cite{braccini} reported on the experimental study of resonant
states in $\gamma\gamma$ collisions at LEP. Quite interestingly, an
$e^+e^-$ collider at high energy can be turned into a powerful tool
for performing a hadron collider-like scan of the various states which
can be produced in photon-photon collisions. In particular, it has been
possible - among other results - to set limits on glueball
candidates and to perform analyses of the charmonium states.

Vittorio Del Duca~\cite{delduca} also reported about photon collision 
studies at
LEP. In this case the photons are taken virtual, so as to set a scale
hard enough for perturbative QCD to be applicable. The process
under consideration is 
\begin{equation}
\begin{array}{rcl}
e^+ + e^- & \longrightarrow & e^+ + e^- + \underbrace{\gamma^* + \gamma^*} \\
 &  & \phantom{e^+ + e^- + \gamma^*\:}\bentarrow {\rm hadrons}
\end{array}
\label{processee}
\end{equation}
The speaker and his collaborators have evaluated next-to-leading (NLO) QCD
corrections for this process, so as to provide a more
reliable theoretical prediction, and therefore a better 
benchmark against which to compare the experimental
results. In this way it will be easier to understand if signals of ``BFKL
dynamics'' are visible in the data.
They have found that NLO corrections can be sizeable and should be
properly included, but the data are still too sketchy to allow a
conclusion about BFKL to be drawn.

A difficult and obscure argument (but not because of this less
important and interesting) was discussed by Lorenzo Vitale~\cite{vitale}, 
namely colour reconnections and Bose-Einstein Correlations. These
effects are due to the fact that when we have many coloured particle in
the final state, like four quarks from $W$ boson decays at LEP2,
phenomena related to the quantum uncertainty principle can take place.
For instance, since the $W$ decay time is much shorter than the
typical hadronization time, we can have a superposition of the two $W$'s
hadronization regions. Hence we can no longer assume they hadronize
independently. For all the details the reader is of course addressed to
Vitale's proceedings. Suffice here to say that the conclusion of the
extremely complex experimental analyses performed is that the variables
studied so far do not have the necessary sensitivity to test the
various models for colour reconnections  proposed, and no significant
effect has been observed. As far as Bose-Einstein Correlations (BEC) are
concerned, there is evidence for the existence of the intra-BEC
(i.e. correlations within the decay products of a single $W$) kind, but no
signal for the inter-BEC one.

The topics I have summarized so far are somewhat unrelated with each
other. They show how varied the studies of QCD can be, and how many
spots worth studying always remain even when the main frontier of the
research moves forward. I now wish instead to describe in some more
detail the topics I consider to represent the mainstream of the westward
path to the gold mines. 

But first, we must of course carefully make sure we are on the correct
path: G\"unther Dissertori~\cite{dissertori}  presented an extensive
list of experimental measurements and checks on the strong coupling
$\alpha_s$. It plays of course a very important role and can
be measured in many different reactions studied at LEP: from very
inclusive ones, like the total hadron production rate, to more exclusive
ones like the various shape variables. Daniele
Bonacorsi~\cite{bonacorsi} also addressed the issue of testing the
agreement between theory and experiment, showing comparisons of
many distributions for multi-hadronic final states measured at LEP with 
the theoretical predictions, in many cases produced by MonteCarlo 
event generators.

In both cases the comparisons are very successful. Hadronic
distributions can be matched to a high degree of accuracy by MonteCarlo
predictions. While it is true that MonteCarlo  models are
usually tuned to (a subset of) the data themselves, the widespread
agreement does however ensure overall consistency of the theoretical
picture. Moreover the precision of the data is good enough 
to exclude some of the more simplified models, in favour of the
ones which include finer details of QCD like, for instance, colour coherence.

The various determinations of $\alpha_s$ extracted from LEP measurements
also show a high degree of internal consistency and good agreement with
the world average from other experiments. All in all, LEP has brought
about a sizeable improvement in the accuracy with which we know
$\alpha_s$, the single largest error on the final measurement being now 
due to the theoretical uncertainty. Indeed, we have for instance
$\alpha_s(M_Z) = 0.1181 \pm 0.0007~({\rm expt}) \pm 0.003~({\rm theo})$
from $\tau$ decay measurements and 
$\alpha_s(M_Z) = 0.1181 \pm 0.0036~({\rm expt}) \pm 0.0052~({\rm theo})$
from shape variables analyses. Both of these results compare very well with
the world average, at present $\alpha_s(M_Z) = 0.1184 \pm 0.0031$, and
show how big a contribution LEP has given to this determination.

In all these measurements and comparisons, two main obstacles have to
be overcome to obtain accurate and meaningful tests. Since we are
always comparing an experimental result to a perturbative series in
$\alpha_s$, it is of course essential to make sure that this series
gives a reliable account of the ``true'' theoretical prediction. Our
result may instead be approximate on two counts: the series might be
poorly converging, and non-perturbative contributions might
have been neglected. These are exactly the two problems that have been
tackled in the talks I am now going to summarize.

Massimilano Grazzini~\cite{grazzini} and Andrea Banfi~\cite{banfi}
reviewed the two items which I consider to represent the frontier of
today's QCD. Albeit distinct, these two issues represent the two ways we are
trying to open our road to a more comprehensive understanding of QCD and
of its non-perturbative region.

On one hand, it is of course very important to have the perturbative
series under good control. The more exclusive an observable is, the more
one can learn from it. But the price to pay is that theoretical
calculations become usually more difficult. Until a few years ago
next-to-leading calculations (i.e. usually one-loop)  were the state of
the art. In the last couple of years, however, a great deal of progress
has been made towards next-to-next-to-leading (NNLO) predictions. They
are particularly important because, if a reliable result in
perturbation theory starts being available at NLO, a reliable estimate
of the error of the prediction can only be achieved at NNLO. The basic
ingredients of a NNLO calculation are  $~i)$ NNLO parton distributions;
$~ii)$ two-loop amplitudes; $~iii)$ knowledge of the IR behaviour of
tree-level and one-loop amplitudes at ${\cal O}(\alpha_s^2)$. Most of these
are now becoming to be available. The talk of Grazzini gives a full
reference list and describes some of the first applications of NNLO
technology.

\begin{figure}[t]
\begin{center}
\epsfig{file=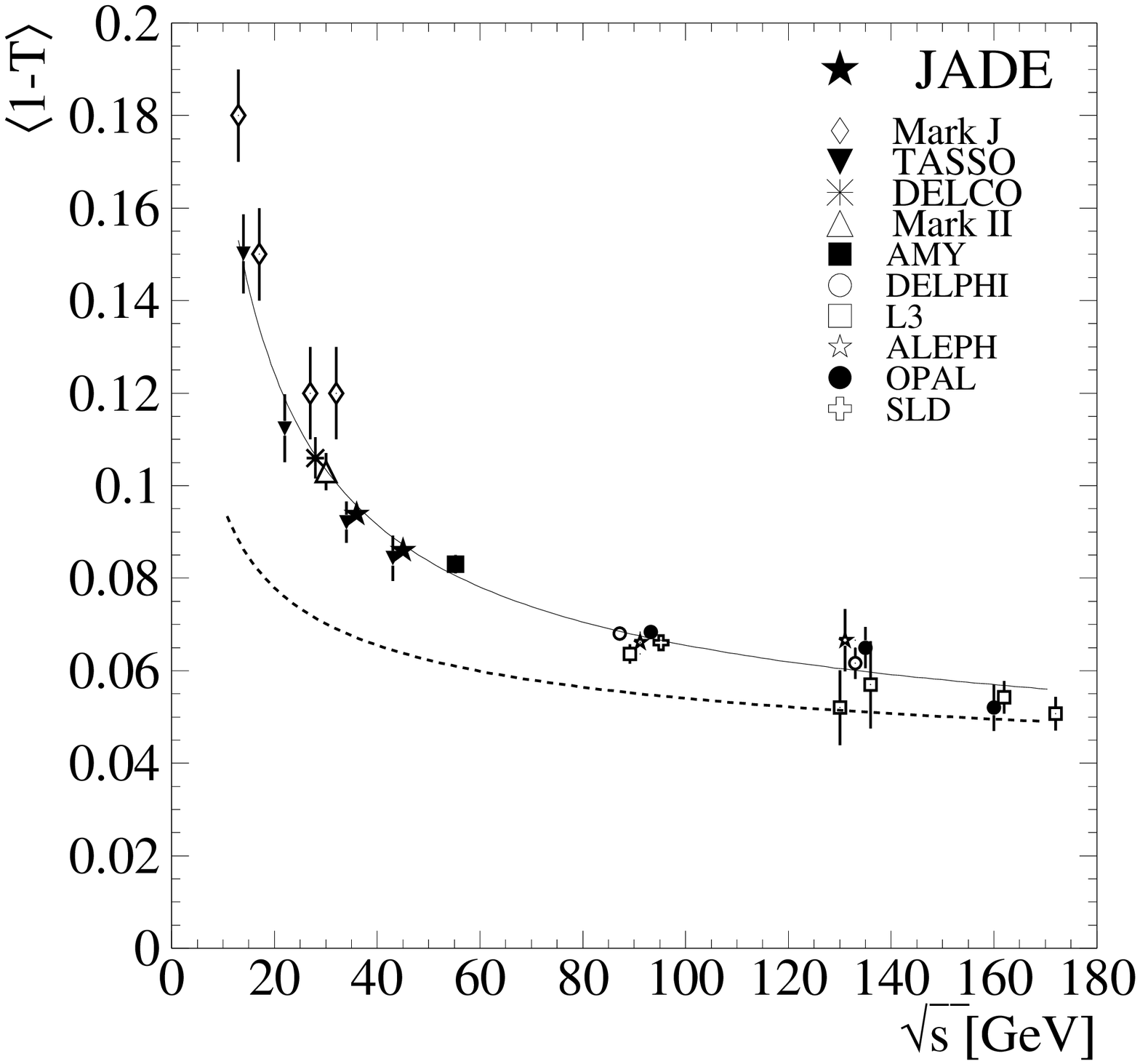,width=7.7cm}
\epsfig{file=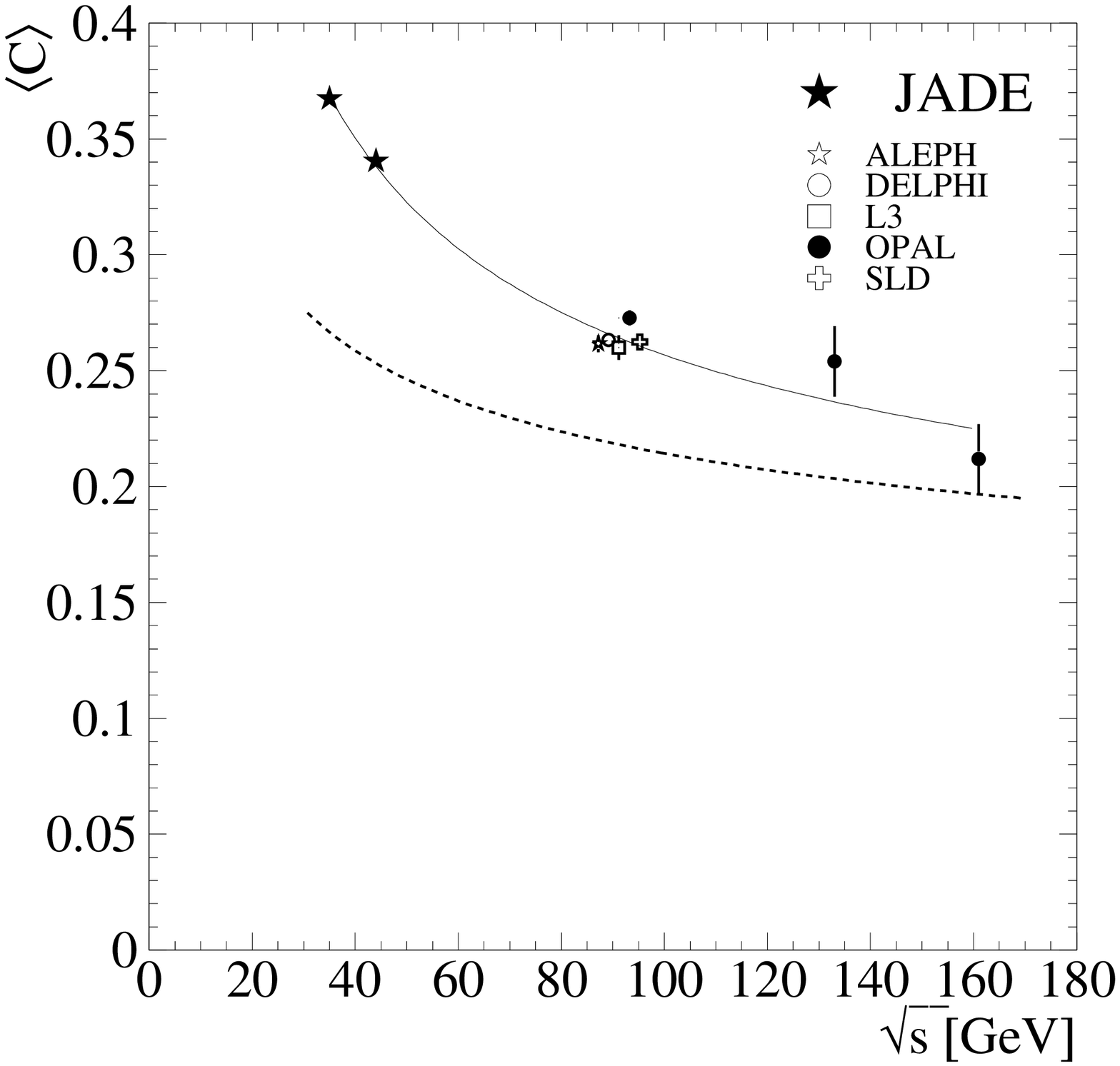,width=7.7cm}\hfill
\caption{\label{fig:thrust}\small Experimental results for the average
value of  $1 -$Thrust and the $C$-parameter, 
compared to a purely perturbative prediction (dashed
line) and to a calculation also including a power correction term (solid
line), see eq.~\protect\ref{eq:thrust}.}
\end{center}
\end{figure}

On the other hand, it is easily understandable how useless it can be to push
to very high accuracy the calculation of a perturbative series, when much
more is left which perturbation theory cannot account for. Andrea
Banfi started by making clear how important non-perturbative power
corrections can be for phenomenology, by showing the plots in
Fig.~\ref{fig:thrust}. In this figure we can see experimental results for
the average value of $1 -$Thrust and the $C$-parameter, as a function 
of the centre-of-mass
energy $\sqrt{s}$. One can clearly see the perturbative prediction,
given by the dashed line, to fail to describe the data. On the other
hand, the inclusion of a power correction of the form 
\begin{equation}
\langle 1- T\rangle = \langle 1-T\rangle_{\rm PT} + \frac{C_T}{\sqrt{s}},
\label{eq:thrust}
\end{equation}
with $C_T \sim 1$~GeV, clearly fits very well the data points. Very
similar plots can be obtained for other variables like, for
instance, the $C$-parameter and the heavy-jet mass. An interesting
aspect is that, say, 
the $C$-parameter results can be fitted with a coefficient
$C_C \sim 4$~GeV, and the ratio $C_C/C_T$ can be correctly predicted 
within a universality hypothesis for $1/Q$ power corrections.
The fact that such a hypothesis appears to work, at least
at this accuracy level, opens the way for a theoretically rigorous 
treatment, which tries to explain many observables within a unified
framework.
One way of parametrizing this universal ``building block'' is
by defining a universal coupling in the infrared region via dispersion 
relations. 
The universality hypothesis can be seen to have some shortcomings, 
and can be somehow extended by introducing a ``shape function'',
that can be used to model non-perturbative effects on many different 
distributions, and not only mean values. 

The important point I wish to stress once more is that these
treatments of power corrections are not purely phenomenological, but
rather try to stick to rigorous theory as much as possible. The
approach is usually that of looking for process independent
behaviour,  and only as a last resort parametrize incalculable parts
in terms of as few as possible measured parameters. The ultimate hope,
of course, is to be eventually able to calculate also these quantities
from first principles.

I wish to conclude with a quote from Paolo Di Vecchia at Moriond QCD
1995. Paolo concluded his talk on the applications of the Seiberg-Witten 
duality approach to the non-perturbative region by saying (I quote from
memory):
\begin{center}
\begin{minipage}{14cm}
\sc
There are two ways of studying non-perturbative QCD:\\
There are people who study QCD-inspired models;\\
And there are people who study theories different from QCD!
\end{minipage}
\end{center}

In that context he was - I think - cautioning that results were indeed
being obtained, but in theories that were not the real QCD.

In the present context, I wish
people will more and more concentrate on trying to provide theoretically sound
solutions for the non-perturbative problems in real QCD, rather than
being tempted to make use of {\sl ad hoc} phenomenological models and
shortcuts.

LEP, and its huge collection of high precision data, has certainly
played a pivotal role in allowing QCD and its phenomenology to make the
transition from a tentative strong interaction theory to a well defined
and widely accepted environment, where precision calculations and
comparisons are possible. Equally accurate data will be very useful in
the future to test the solutions for the non-perturbative region we may
come up with. Since no new machine capable of producing them will be 
available any time soon, it is imperative we try to save the LEP results
for future use, in a form where the bias from present-day knowledge of
non-perturbative processes is as small as possible.

\vspace{1cm}
\noindent
{\bf Acknowledgments.} I wish to thank the Organizers for the
invitation and Fabrizio Fabbri, who shared the convening of this 
parallel session and explained to me
the details of many experimental results. I am also mostly grateful to
all the speakers for their efforts and availability. Needless to say, the
responsibility for how this Summary was framed only rests with me.

\end{document}